# Local manifestations of thickness dependent topology and axion edge state in topological magnet MnBi$_2$Te$_4$

Felix Lüpke[1,2,*], Anh D. Pham[1], Yi-Fan Zhao[3], Ling-Jie Zhou[3], Wenchang Lu[4,5], Emil Briggs[4], Jerzy Bernholc[4,5], Marek Kolmer[1,**], Wonhee Ko[1], Cui-Zu Chang[3], Panchapakesan Ganesh[1,†], An-Ping Li[1,††]

[1]Center for Nanophase Materials Sciences, Oak Ridge National Laboratory, Oak Ridge, Tennessee 37831, USA

[2]Department of Materials Science and Engineering, University of Tennessee, Knoxville, Tennessee 37916, USA

[3]Department of Physics, The Pennsylvania State University, University Park, Pennsylvania 16802, USA

[4]Department of Physics, North Carolina State University, Raleigh, North Carolina 27695, USA

[5]Computational Sciences and Engineering Division, Oak Ridge National Laboratory, Oak Ridge, Tennessee 37916, USA

*Present address: Peter Grünberg Institut (PGI-3), Forschungszentrum Jülich, 52425 Jülich, Germany

**Present address: Ames Laboratory - U.S. Department of Energy, Ames, Iowa 50011, USA.

† ganeshp@ornl.gov

†† apli@ornl.gov

**The interplay of non-trivial band topology and magnetism gives rise to a series of exotic quantum phenomena, such as the emergent quantum anomalous Hall (QAH) effect and topological magnetoelectric effect. Many of these quantum phenomena have local manifestations when the global symmetry is broken. Here, we report local signatures of the thickness dependent topology in intrinsic magnetic topological insulator MnBi$_2$Te$_4$ (MBT), using scanning tunneling microscopy and spectroscopy on molecular beam epitaxy grown MBT thin films. A thickness-dependent band gap with an oscillatory feature is revealed, which we reproduce with theoretical calculations. Our theoretical results indicate a topological quantum phase transition beyond a film thickness of one monolayer, with alternating QAH and axion insulating states for even and odd layers, respectively. At an even-odd layer step, a localized gapped electronic state is observed, in agreement with an axion insulator edge state that results from a phase transition across the step. The demonstration of thickness-dependent topological properties highlights the role of nanoscale control over novel quantum states, reinforcing the necessity of thin film technology in quantum information science applications.**

Topological insulators (TIs), such as $Bi_2Te_3$, have an inverted band gap which is manifested as topologically protected Dirac-like surface states[1-4]. By introducing magnetism into a topological insulator, the time-reversal-symmetry of the system is broken and thus a magnetic exchange gap opens at its Dirac point[5]. When the chemical potential is tuned into this magnetic exchange gap, the quantum anomalous Hall (QAH) effect with dissipation-less chiral edge transport appears. The QAH effect has been experimentally realized in magnetically doped $Bi_{2-x}Sb_xTe_3$ thin films[6,7]. However, it has become clear that the random distribution of dopants (e.g. Cr or V) can lead to inhomogeneous magnetic exchange gaps[8] and, more importantly for thin films, local chemical potential fluctuations[9], which undermine the promises of the unique electronic properties in potential applications.

Recently, it was shown that stoichiometric $MnBi_2Te_4$ (MBT) provides an alternative way to realize a magnetic TI. In contrast to a TI with randomly distributed magnetic dopants, MBT has an ordered crystal structure in which the magnetic moments are arranged in layers with an A-type antiferromagnetic (AFM) ordering[10-14]. As a result, MBT promises a bigger, more uniform magnetic exchange gap and chemical potential than magnetically doped TIs[13,15,16]. More interestingly, MBT thin films exhibits alternating topological phases with its thickness. In MBT thin films with an odd number of layers, a zero magnetic field QAH state has been theoretically predicted[11,17] and experimentally demonstrated in transport measurements[16]. In contrast, the axion insulator state was claimed to be realized in thin films with an even number of layers[18]. However, while the local electronic structure of bulk MBT crystals has been studied in detail with scanning probe techniques[19-22], the thickness-dependent variations of electronic properties of the MBT thin films remain elusive. Especially, how a symmetry breaking introduced by thickness changes will affect the quantum states in MBT remains to be explored.

Here, we report the structural and spectroscopic characterization of epitaxial MBT thin films by scanning tunneling microscopy/spectroscopy (STM/S) in combination with first-principles theoretical modeling to examine the thickness-dependent electronic properties underpinning the reported QAH and axion insulator states. Moreover, we report the observation of edge states as a local manifestation of symmetry breaking, which indicates the realization of the axion insulator state and offers a glimpse at the magnetic structure of the MBT film.

MBT films with a nominal thickness $t$ of several septuple layers (SL) are grown on a bilayer graphene terminated 6H-SiC(0001) substrate by molecular beam epitaxy. The growth procedure consists of alternating deposition of $Bi_2Te_3$ and MnTe layers followed by annealing as reported in a prior study[15]. After the growth, samples are capped with ~10 nm of Te and transferred to a combined scanning electron microscope (SEM)/STM chamber (Omicron LT Nanoprobe) where the Te capping layer is removed by annealing the sample at ~ 250°C in UHV for 60 min. The subsequent STM/S measurements are performed at 4.6 K.

Scanning electron microscopy measurements of the de-capped sample show island growth (Fig. 1a). Large area STM scans (Fig. 1b) reveal MBT island thicknesses in multiples of $(13 \pm 0.3)$ Å which corresponds to a thickness of $t = 1$ SL. Atomic resolution scans on the islands (Fig. 1c) show a hexagonal lattice with lattice constant $a = 4.3$ Å, consistent with the Te plane at the MBT surface (Fig. 1d) and in agreement with measurements on MBT bulk crystals[22]. Furthermore, low-energy electron diffraction (LEED) measurements on the thin film (Fig. 1e) show a streaky diffraction pattern, corresponding to $a = 4.3$ Å, in addition to the well-known graphene/SiC substrate spots[23]. The MBT spots indicate a preferred rotational alignment of the MBT with the graphene orientation. From the width of the MBT spots, we determine the alignment of the MBT islands with respect to the substrate to have a standard deviation of ∼ 6.4° (see supplemental material).

Systematic scanning tunneling spectroscopy study of the MBT islands with thicknesses ranging from 1 SL to 6 SL (Fig. 2a) show overall similarity to spectra taken on bulk MBT crystals[22] but are shifted in energy and varying in the gap sizes. Independent of the layer thickness, we find that the conduction band edges of the films are located near the Fermi energy $E_F$. This band shift can be attributed to the same lattice defects that render bulk MBT electron doped, particularly Te vacancies and anti-site defects[21,24]. This observation is also in agreement with other vdW thin films grown on Gr/SiC substrates[25]. Our results further show a systematic change of the MBT island band gap sizes as function of their layer thickness. The extracted band gaps (Fig. 2b) show a sharp decrease in gap size from 1 SL (∼ 321 meV) to 2 SL (∼ 152 meV), followed by a gradual increase of the band gap up to ∼ 200 meV for 6 SL (the thickest film studied here). Interestingly, odd layers show slightly smaller gaps while even layers show slightly bigger gaps, with an oscillatory variation with layer thickness. The sizes of the observed gaps generally agrees well with recent photoemission experiments[26], but are significantly larger than theoretically predicted gaps[17].

To capture the experimental trend and obtain a quantitative comparison of the measured band gaps, we performed density functional theory (DFT) calculations with improved treatment of non-local exchange and correlation effects, which allow us to explore the sensitivity of band gaps to interlayer separation in thin-films as well as local magnetic moments. A summary of our calculations is shown in Fig. 2b alongside the experimental data. Using Perdew-Burke-Ernzerhof (PBE) functionals and a Hubbard U of 4 eV as reported by neutron experiments[13], our theoretical calculations qualitatively reproduce the trend of experiment (for details on the calculations see supplemental material). Interestingly, even when using the same DFT-D3 type dispersion correction and a Hubbard U of 4 eV[27,28], we find that the interlayer distances obtained from PBE are generally larger than those obtained from the PBESol functional by ∼0.5 Å, with the PBE interlayer separation tending towards the bulk value of 13.6 Å[29] for 6 SL (Fig. S2). However, for thicknesses of 3 SL to 4 SL, the PBEsol+U functional[30], which is an improved functional for solids, predicts the interlayer distance to be close to 13.1 Å, which is in better agreement with our experimentally measured layer thickness of $(13.0 \pm 0.3)$ Å. Consequently, due to different vdW interlayer distances, there are quantitative differences in the band gaps as function of thickness. Overall, we find that for thicknesses of 3 SL and 4 SL, the structures with smaller interlayer separation, obtained from PBESol+U, show a larger gap than the ones obtained from PBE+U. Using a real-space code (RMG-DFT[31-33]) with harder pseudopotentials, we find a similar trend for the band gaps as from our plane-wave VASP calculations. For both, PBE+U and PBESol+U, even layers show a larger gap than odd layers, due to the antiferromagnetic ordering which results in uncompensated moments in odd layers. This even-odd oscillatory behavior on the bandgap is consistent with our experiments and is reduced with increasing layer-thickness as expected in the thermodynamic limit. Quantitatively though, PBE/PBESol+U functionals still significantly underestimate the gaps observed in the experiments.

| $t$ (SL) | 1 | 2 | 3 | 4 | 5 | 6 |
|---|---|---|---|---|---|---|
| $\Delta$ (meV) | 323 | 147 | 157 | 174 | 175 | 198 |
| $\alpha$ (%) | 10 | 10 | 35 | 37 | 25 | 43 |
| $\mu_B$/Mn | 4.27 | 4.26 | 4.47 | 4.48 | 4.41 | 4.51 |

Table 1: Band gaps $\Delta$, exact-exchange values $\alpha$ and moments per Mn atom as function of layer thickness $t$ determined by fit of HSE06 hybrid functional calculations to the experimental data.

To resolve this discrepancy, we resort to the more computationally expensive Heyd-Scuseria-Ernzerhof type hybrid functionals (HSE06)[34-36], which includes a varying degree of exact-exchange interaction to the total potential, to capture non-local effects in the correlation of electrons with the same spin. From calculations with the hybrid functional approach, we find that adding the exact-exchange generally results in larger gaps than obtained from PBE+U and PSESol+U (see supplemental material for details). By varying the exact-exchange, we fit the band gaps obtained from PBESol+U+D3 to the experimental gap data, resulting in the values listed in Table 1. These results directly indicate that the thinner films (1 SL and 2 SL) have ~0.2 $\mu_B$ lower local moment per Mn atom compared to the thicker films (3 SL to 6 SL), which are close to the bulk value (~4.51 $\mu_B$ per Mn), as detailed in the supplemental material. Even layers in general have similar sized or higher moments per Mn atom, than the thinner odd layers, thus explaining the larger gaps observed on the even layers compared to the odd layer due to the overall antiferromagnetic ordering.

Independent of the employed functionals, our calculations show a sharp decrease of the band gap from 1 SL to 2 SL followed by the opening of an inverted gap for ≥ 2 SL. This inverted band gives rise to QAH and axion insulator states in odd and even layers, respectively. We calculated the resulting edge states in a 3 SL and 4 SL film (Fig. 3). For a 3 SL film, the uncompensated magnetic moments result in the breaking of time reversal symmetry, which is manifested as a topologically protected gapless edge states, namely, the QAH edge state. In contrast, for a 4 SL film, the combination of preserved time-reversal-symmetry and half translation symmetry ($T\Theta_{1/2}$) results in an edge state that is not topologically protected and for which our DFT calculations predict a gap of 20 meV. We explore the latter edge state experimentally by studying a 4 SL – 3 SL step edge (Fig. 4a). STS measurements (Fig. 4b) across the step edge show the presence of an increased $dI/dV$ signal on the 4 SL terrace, next to the topographic position of the step edge. The spectroscopic signature of this edge state is an increased density of states in the energy range $-0.35$ eV $\leq$ E $\leq$ 0 eV, i.e. extending into the $\Delta \approx 175$ meV band gap observed on the 4 SL terrace (Fig. 4c). In qualitative agreement with our calculations, this edge state has a gap of $\Delta \approx 50$ meV which is located at the Fermi energy. In differential conductance maps corresponding to the edge state energy ($-300$ meV), we find that the edge state extends along the topographic edge of the 4 SL terrace (Fig. 4d), similar to the case of, e.g., the quantum spin Hall edge state in WTe$_2$[25,37]. The observation of the gapped edge state in the 4 SL film indicates an AFM alignment between the layers such that the individual layer's magnetic moments are compensated, because if the film had a net magnetic moment we would expect a gapless QAH edge state based on theory. In combination with the remarkable agreement between our film thickness-dependent theoretical calculations of the electronic structure with our experimental STS data, we therefore conclude that the observed edge state feature is the manifestation of the axion insulator edge state.

In conclusion, we have demonstrated the local manifestation of QAH and axion insulator phases in MBT thin films with varying thickness. The observation of a gapped edge state at an even-odd layer step edge confirms the existence of the axion insulator phase with net compensated magnetic moments. In contrast, an odd-even step edge, e.g. a 3 SL – 2 SL step, is expected to host the topologically protected metallic QAH edge-state. However, owing to a dominant presence of even layer thickness islands in our samples we have not experimentally observed such a QAH edge-states, yet. We take the observation of predominantly even layer thicknesses, i.e., layers with compensated net magnetic moment, as an indication that MBT films with even layer thicknesses are energetically favored over odd layer thicknesses during the film growth. This finding thus motivates the optimization of film growth to promote odd layer coverages in order to realize the QAH edge state at odd-even step edges. In combination with *s*-wave superconductivity, the QAH and axion state in MBT thin films provide a route towards the realization of Majorana zero modes for topological quantum computing[38].

From a theoretical point of view, the improved agreement of our calculations with experiments, by variation of the exact exchange for different thicknesses, suggests a complicated interplay between electronic correlations, magnetism, and hybridization. This makes MBT an interesting benchmark system to validate a variety of theoretical methods beyond the conventional DFT approach such as the dynamic mean field theory and quantum Monte Carlo methods[39]. Furthermore, while plane-wave DFT-methods do not allow explicit simulations of step-edges, we demonstrate that DFT methods using multi-resolution real-space grids are suitable for large-scale simulations[31-33], which agree well with plane-wave codes and enable calculations of edge states for arbitrary film thicknesses in future studies.

**Acknowledgements**

This research was conducted at the Center for Nanophase Materials Sciences, which is a DOE Office of Science User Facility. F.L. acknowledges funding from the Alexander von Humboldt foundation through a Feodor Lynen postdoctoral fellowship. A.D.P. was financially supported by the Oak Ridge National Laboratory's Laboratory Directed Research and Development project (Project ID 7448, PI: P.G.). The VASP calculations used resources of the National Energy Research Scientific Computing Center (NERSC), a U.S. Department of Energy Office of Science User Facility operated under Contract No. DE-AC02-05CH11231. All computations using Wannier90 code used resources of the Compute and Data Environment for Science (CADES) at the Oak Ridge National Laboratory, which is supported by the Office of Science of the U.S. Department of Energy under Contract No. DE-AC05-00OR22725. The development of RMG was funded by the Department of Energy Exascale Computing Project and the National Science Foundation grant OAC-1740309. RMG based computations used resources of the Oak Ridge Leadership Computing Facility at the Oak Ridge National Laboratory, which is supported by the Office of Science of the U.S. Department of Energy under Contract No. DE-AC05-00OR22725. The film growth done at Penn State is supported by the Gordon and Betty Moore Foundation's EPiQS Initiative (Grant GBMF9063 to C.Z.C.) and ARO Young Investigator Program Award (W911NF1810198). A portion of the research (A.-P.L.) is supported by the U.S. Department of Energy, Office of Science, National Quantum Information Science Research Centers.

**Figures and Figure captions:**

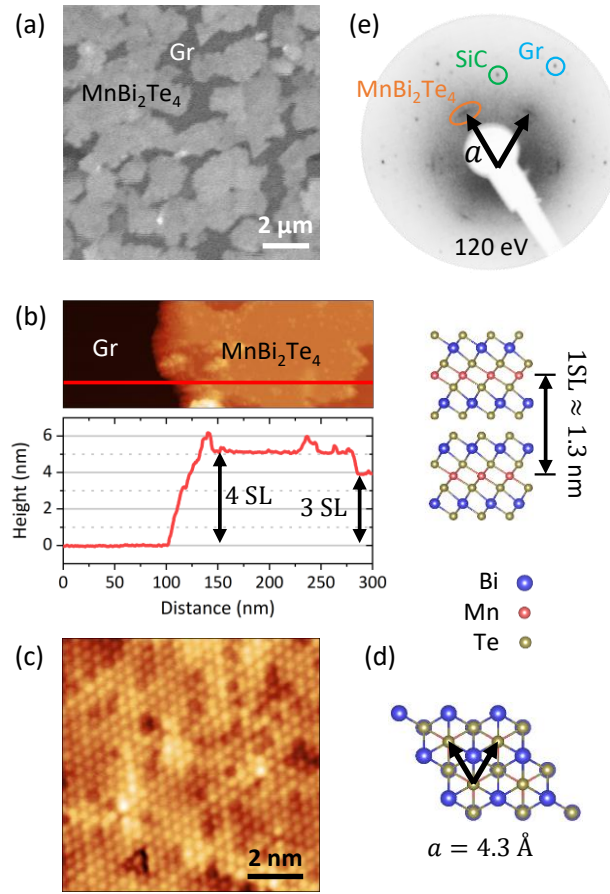

**Figure 1. Epitaxial MnBi$_2$Te$_4$ thin films on bilayer graphene terminated SiC.** (a) Scanning electron micrograph of the sample showing an island growth mode. (b) Scanning tunneling micrograph of a MnBi$_2$Te$_4$ island with different thicknesses on the substrate (top). Height profile along the indicated line and atomic model of MnBi$_2$Te$_4$ with indicated layer-to-layer distance (bottom). (c) Atomic resolution tunneling image acquired on an island showing a hexagonal lattice with lattice constant $a = 4.3$ Å. (d) Top view of the MnBi$_2$Te$_4$ lattice with indicated lattice vectors corresponding to the spacing of the topmost Te layer. (e) Low-energy electron diffraction pattern showing streaks corresponding to $a = 4.3$ Å, indicating a variation in the island orientation around a preferred alignment with the graphene lattice.

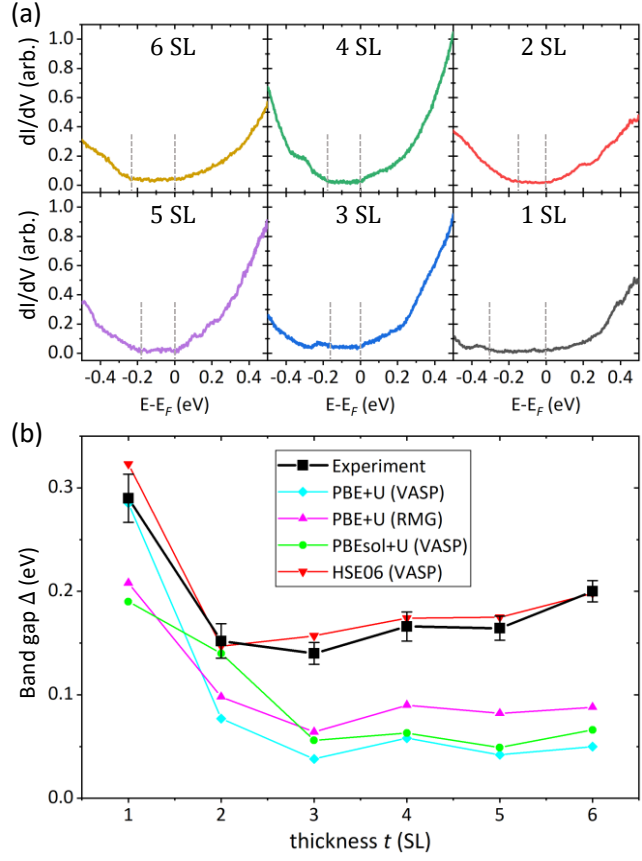

**Figure 2. Thickness-dependent electronic structure of MnBi$_2$Te$_4$.** (a) Tunneling conductance measured on MnBi$_2$Te$_4$ islands of different thickness with valence and conduction band edges indicated (see supplemental material for details). (b) Thickness-dependent experimental band gaps and comparison to theoretical calculations using different kinds of functionals. Independent of the functional used for the calculations, the band gap is topological (inverted) above $\geq 2$ SL thickness.

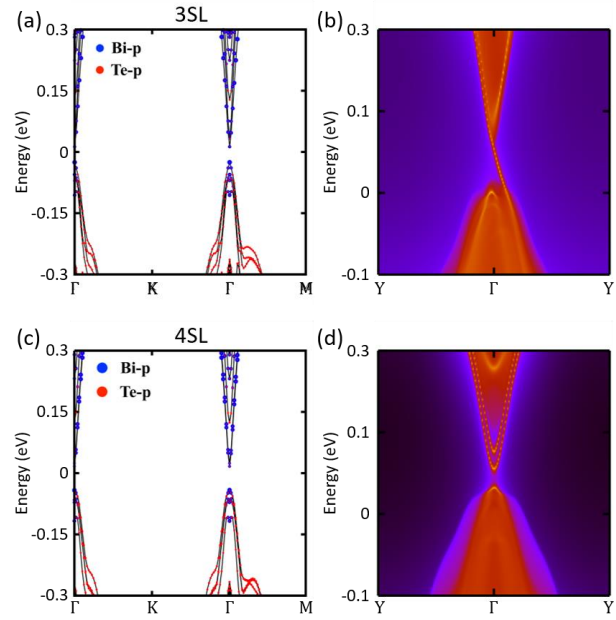

**Figure 3. Calculated band structure and edge states of a 3 SL and 4 SL MnBi$_2$Te$_4$ film.** (a) Bulk band structure of 3 SL film with (b) a gapless QAH edge state which crosses the band gap. (c) Bulk band structure of 4 SL film with (d) a topologically trivial edge state with a gap.

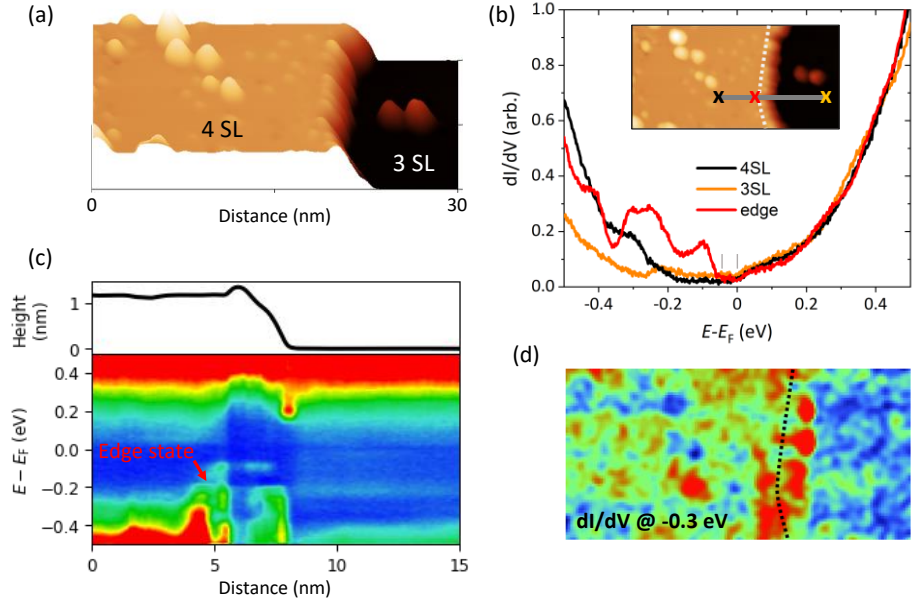

**Figure 4. Observation of axion edge states.** (a) Topography of a 1 SL high step edge. (b) Tunneling conductance at different locations indicated in the inset. Away from the edge, the spectra on the terraces correspond to the spectra shown in Fig. 2. Near the position of the edge (indicated as dashed line in the inset) an additional density of states is observed indicating the presence of an edge state. (c) Tunneling spectra taken along the line indicated in the inset in (b) aligned with the height profile (top) along the same line. The increased density of states of the edge state feature is visible on top of the flat terrace within vicinity to the topographic edge. Stabilization parameters: $V_{\text{sample}} = 0.5$ V, $I_t = 100$ pA. (d) Intensity map of the tunneling conductance at $V_{\text{sample}} = -0.3$ eV showing the spectral feature of the edge state being present continuously along the edge.

*Supplemental Material*

**LEED pattern analysis**

To determine the variation in the MBT island orientation, we estimate the width of the corresponding peak in the LEED pattern from a cross section along a circular path (Fig. S1). A Gaussian fit $y = y_0 + Ae^{-\frac{1}{2}\left(\frac{x-x_c}{w}\right)^2}$ to the MBT peak results in a width of $w \approx 6.7°$. For comparison, the substrate peak width is $w_{sub} \approx 1.9°$. Since the substrate is expected to be perfectly aligned across the LEED beam spot (~1 mm²), we attribute the substrate peak width to be due to instrumentation and estimate the actual angle deviation of the MBT to be $w_{MBT} = \sqrt{w^2 - w_{sub}^2} \approx 6.4°$.

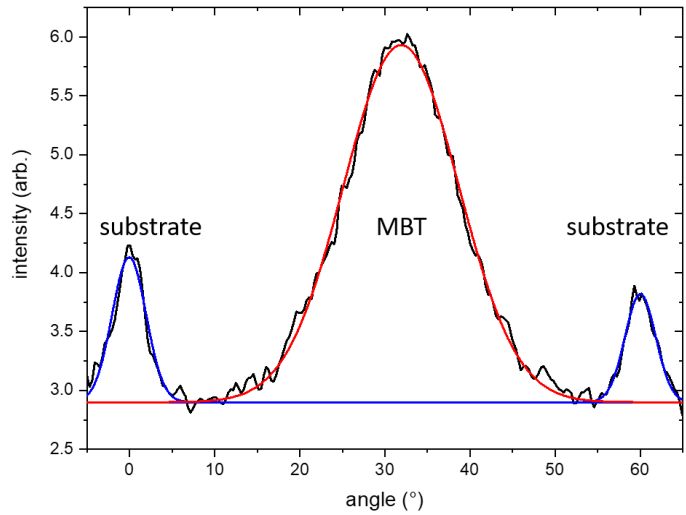

**Figure S1.** MBT spot profile extracted from LEED pattern along circular path with Gaussian fits.

**Details on DFT calculations**

The PBE+U calculations for the thin films were performed using PAW potentials as implemented in the VASP code. A Hubbard $U = 4$ eV[1] was used to describe electrons in Mn's d orbitals to account for electron-electron correlation effects. We set the lattice constants $a = b = 4.36$ Å for the different film thicknesses, consistent with previous theoretical studies of MBT thin films[2,3]. The internal coordinates were relaxed till the forces were down to 0.01 eV/atom, with a plane-wave cut-off energy of 400 eV and a dense $\boldsymbol{k}$-mesh of $9 \times 9 \times 1$ to integrate the Brillouin zone. Relaxation were performed using the exchange-correlation functionals PBE[4] with a DFT-D3 van der Waals correction scheme[5,6]. In addition, calculations were also performed using the PBEsol[7] functional with $U = 4$ eV and a DFT-D3 vdW correction to test the effect of different functionals on the interlayer distance and the band gaps for films at different thickness. For both of the PBE+U and PBEsol+U calculations, we used the potentials Bi_d, Mn_sv and Te for our calculations.

The band gap calculations as well as topological analysis reported in the manuscript were performed after incorporating spin-orbit coupling (SOC). We investigated the topological properties of the films by constructing a tight-binding Hamiltonian using $U = 4$ eV from PBE+U calculations, by projecting the plane-wave basis to the atomic like Mn-$d$ orbitals, Bi-$p$ orbitals and Te $p$-orbitals using the Wannier90 package[8]. The edge states were calculated using the WannierTools code[9].

The Heyd-Scuseria-Ernzerhof (HSE06) type hybrid functional[10-12], has a range-separated Coulomb potential, with a fraction of the exact-exchange added only to the short-range part of the potential, with the long-range part of the Coulomb potential consisting of PBE-exchange, making it somewhat more computationally tractable. Inclusion of exact-exchange while capturing non-local electron-electron interaction more accurately makes the calculations significantly more expensive. As such, in this set of calculations, we used the softer pseudo-potentials Bi, Mn and Te for these calculations due to the memory and time constraints. We first relax the geometries again using 270 eV cutoff with $9 \times 9 \times 1$ using the PBEsol+U ($U = 4$ eV) and the soft potentials to determine the structural ground state. There is minimal difference between the Mn-Mn interlayer distance using the two sets of potentials for PBEsol+U functional. We then utilized these set of structures for our band gap calculations. A Gamma centered $k$-mesh of $3 \times 3 \times 1$ was used to integrate the Brillouin zone. It is possible to vary the fraction of exact-change which we include in the short-range part of the potential.

Typically, Hartree-Fock theory results in larger bandgaps, while local or semi-local density functionals lead to an underestimation of the gaps[7]. But the effect on inverted bandgaps is not so clear, since the size of the inverted-gap is determined by the strength of the SOC and the magnitude of the exact-exchange. In our hybrid calculations, we find that adding (25%) exact-exchange leads to the inverted bandgaps (Fig. S4) becoming closer in agreement to the experiments for the thicker films (Table S1), and generally larger than our gaps from PBE/PSESol+U results. Typically, the bandgap depends on thickness[13], as seen in other (non-SOC) 2D materials, due to a quantum confinement effects – with the bandgap increasing in magnitude as the film thickness becomes comparable to the exciton radius. This indicates that with thickness, due to quantum confinement effects, there will be changes in the dielectric screening of the material and this will effectively change the degree of electron-electron correlation.

Understanding changes in electronic properties with thickness, particularly in the presence of heterogeneities, such as vacancies and anti-site defects (not studied here), requires large-scale simulations, not currently feasible with methods using plane-wave basis. Real-space DFT-methods can perform large-scale simulations for which we developed new capabilities enabling DFT-calculations using the Hubbard-U and SOC in the Real-space MultiGrid (RMG) code v.4.0[14-16]. We further benchmarked the variation of the bandgap with thickness against plane-wave codes. RMG is an open-source DFT-based code that uses real-space grids to represent wave functions, charge densities and the ionic pseudopotentials. It is designed for scalability and performance, using multigrid techniques to accelerate convergence and domain decomposition to parallelize well up to thousands of nodes containing multiple multi-core CPUs and GPU accelerators. The spin orbit coupling effect is included by using fully relativistic pseudopotentials either in the optimized norm-conserving Vanderbilt (ONCV) or ultrasoft format[17-19]. The RMG calculations presented here utilized Hamann's ONCV pseudopotentials[19]. To improve the accuracy, semi-core levels are also included in valence: Mn $3s$ and $3p$, Bi $5d$, and Te $4d$. The grid spacing for the wave functions is 0.22 bohr, which corresponds to 118 Ry cutoff in plane-wave calculations. The corresponding plane-wave cutoff for the electron density and potentials is 475 Ry. For consistency, the same atomic structures from PBE+U calculations were used for VASP and RMG calculations. The RMG calculations were run on up to 96 nodes of ORNL's Summit supercomputer. Each node contains 2 IBM POWER9 CPUs and 6 Nvidia Volta GPUs.

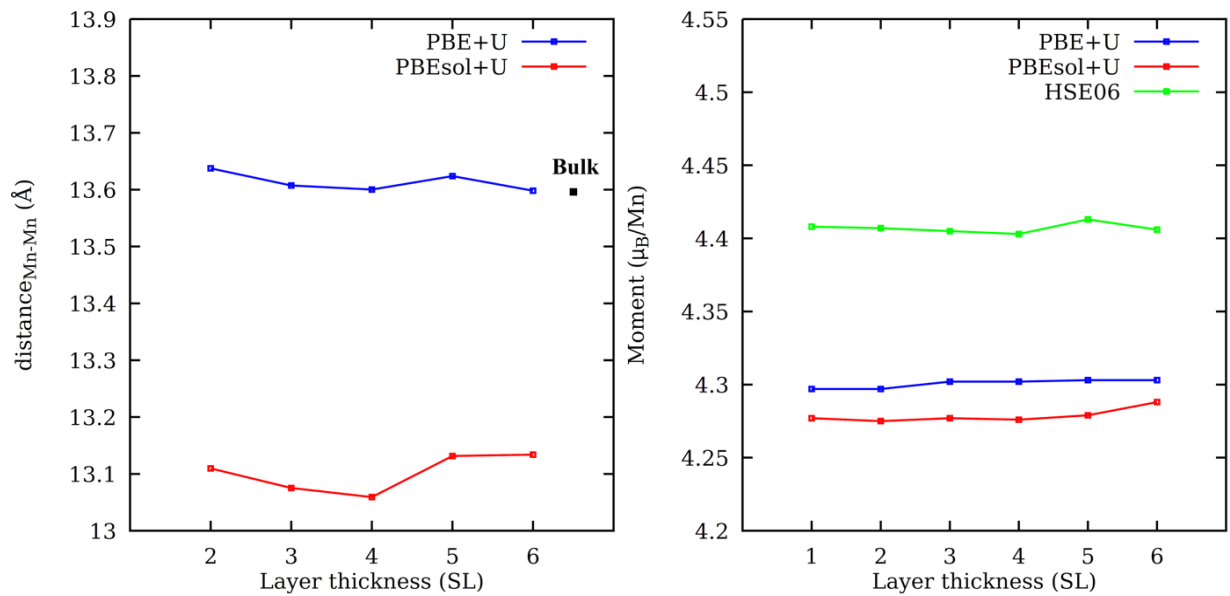

**Figure S2**. Dependence of the interlayer distance and magnetic moment on the different functionals.

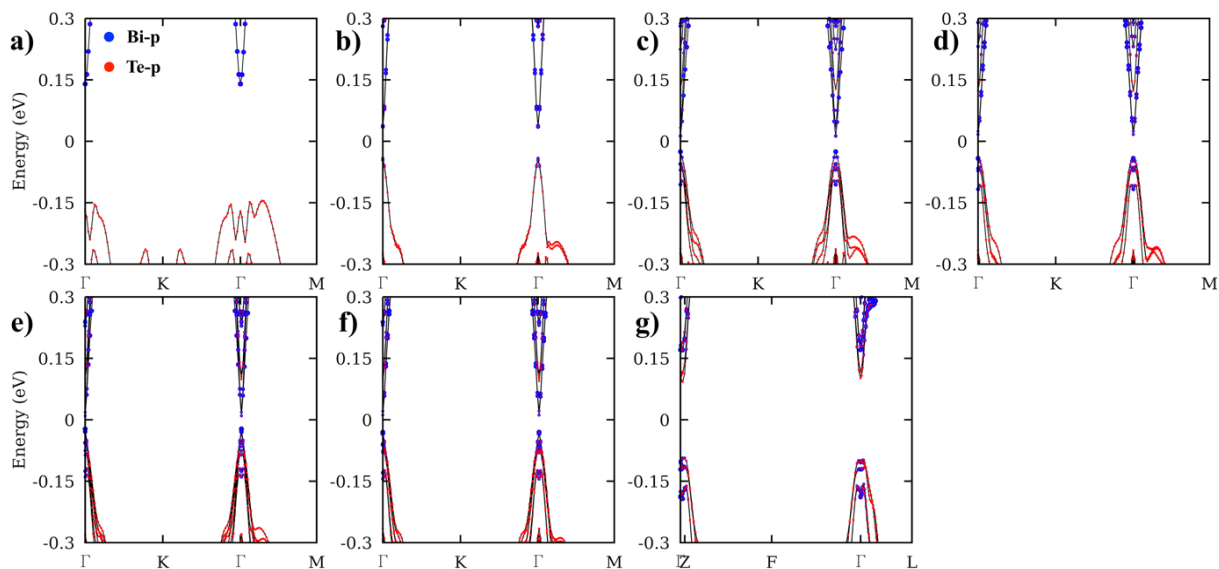

**Figure S3.** Band structure of MBT at different thicknesses calculated with PBE+U ($U = 4$ eV) functional with DFT-D3 correction (and resulting band gaps). (a) 1 SL (285 meV). (b) 2 SL (77 meV). (c) 3 SL (38 meV). (d) 4 SL (58 meV). (e) 5 SL (42 meV). (f) 6 SL (49.8 meV). (g) Bulk (186 meV).

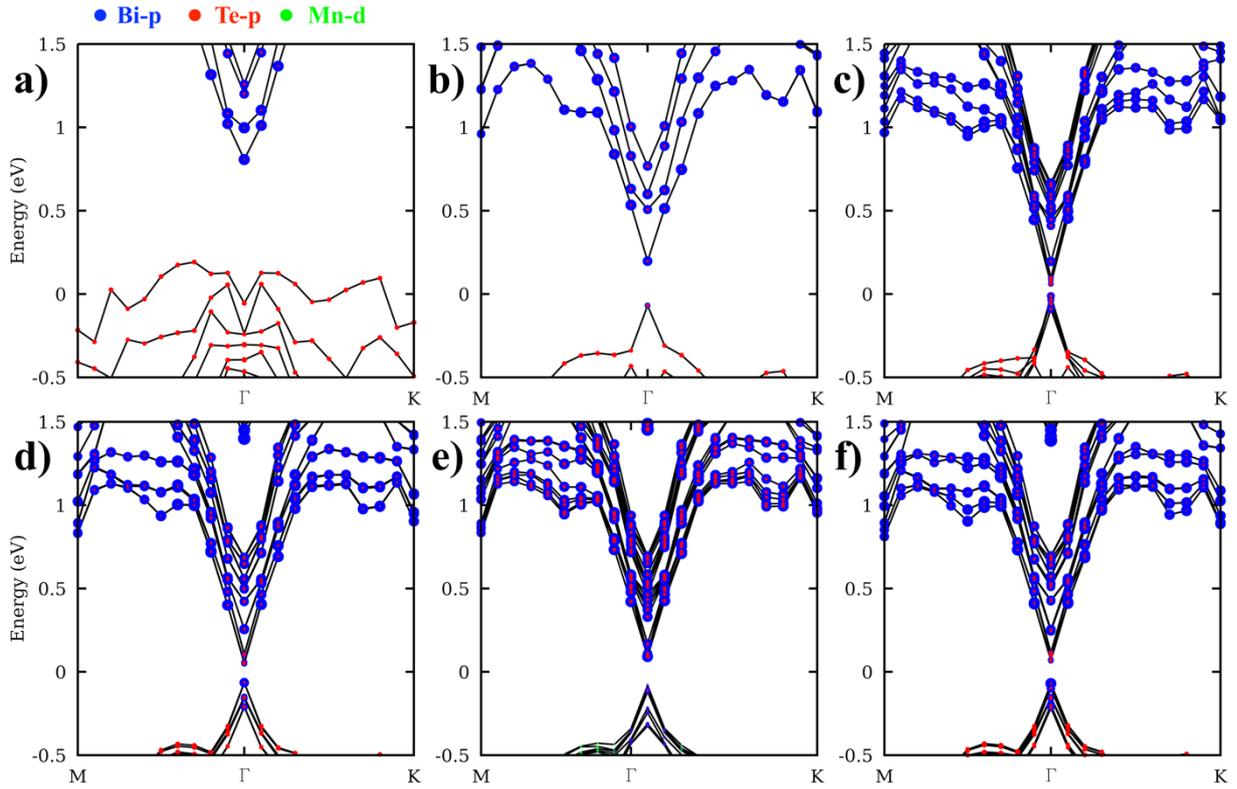

**Figure S4.** Band structure of MBT at different thicknesses calculated with HSE06 with $\alpha = 0.25$ (and resulting band gaps). (a) 1 SL (615 meV). (b) 2 SL (266 meV). (c) 3 SL (65.6 meV). (d) 4 SL (113 meV). (e) 5 SL (175 meV). (f) 6 SL (147 meV). The band gaps are inverted for thicknesses $\geq$ 2 SL).

|      | Exact exchange fraction $\alpha$ | | | | | | | | |
|------|------|------|------|------|------|------|------|------|------|
|      | 0.1  | 0.15 | 0.25 | 0.3  | 0.35 | 0.37 | 0.4  | 0.43 | 0.45 |
| 1 SL | **323** | 415 | 615 |     |     |     |     |     |     |
| 2 SL | **147** |    | 266 |     |     |     |     |     |     |
| 3 SL |     |     | 65.6 | 98.7 | **157** |     | 225 |     |     |
| 4 SL |     |     | 113 | 114 | 155 | **174** | 214 |     |     |
| 5 SL |     |     | **175** |     |     |     |     |     |     |
| 6 SL |     |     | 147 |     |     |     | 153 | **198** | 235 |

**Table S1.** Band gap size (in meV) as function of the exact exchange $\alpha$ at different film thicknesses. The values in bold are plotted in Fig. S5 and Fig. 2 of the main text.

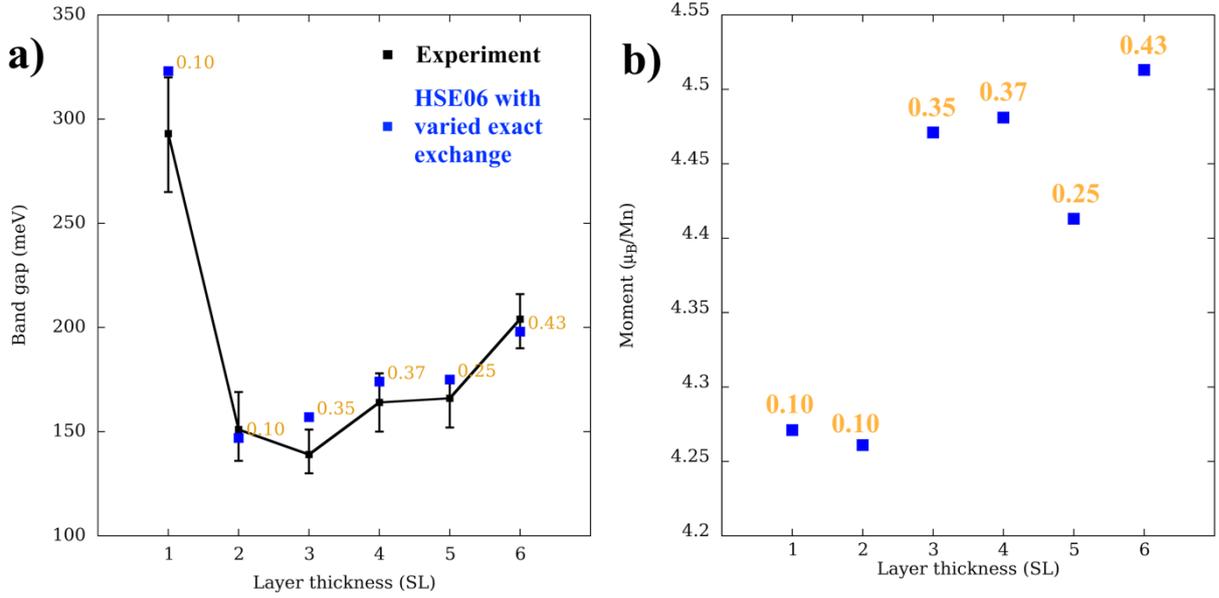

**Figure S5**. Electronic gaps and magnetic properties at different thickness calculated with varied exact exchange fraction $\alpha$. (a) Fit of exact exchange to the experimental gaps. (b) Magnetic moment at the corresponding exact exchange fraction (highlighted values in Table S1).

**Edge state transition between the quantum anomalous Hall state and axion state**

The axion state in the four SL film is shown to be a fragile topological state, which strongly depends on the film magnetism as well as the layer coupling. To demonstrate this effect, we split the total Hamiltonian into one for the topmost SL and one for the bottom 3 SL i.e. $H_{\text{total}} = H_{\text{bottom 3 layers}} + \varepsilon H_{\text{top layer}}$. With no coupling to the top most layer (i.e. $\varepsilon = 0$), the electronic properties of the 4 SL film resemble a 3 SL film which exhibits the quantum anomalous Hall effect with a single edge state (Fig. S6). As we increase the layer coupling, the overall magnetization decreases, resulting in a topological transition with the film gap closing and reopening at $\varepsilon \approx 0.83$. This topological transition is accompanied by the transition from a single chiral edge state to a gapped Dirac cone.

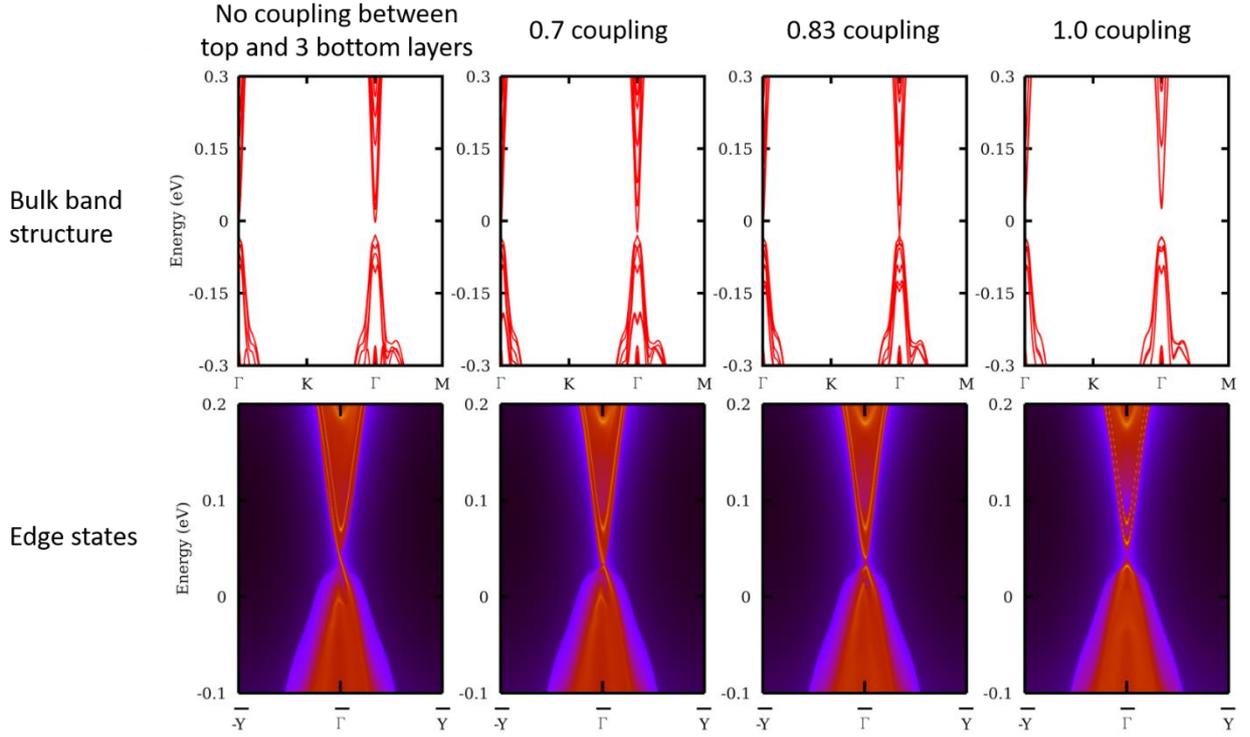

**Figure S6.** Electronic structure of a 3 SL film where the coupling ε between the topmost layer and the underlying three layers is tuned.

**Band gap extraction**

We extract the thickness dependent band gaps from log-plots of the $dI/dV$ spectra by linear extrapolation of the band edges. For each island thickness, we have acquired tunneling spectra on three different locations and have checked the tip spectrum on the Gr/SiC substrate between measurements. The extracted band gaps shown in Fig. 2b of the main text are the average of the three gaps, which we extract for each thickness, and error bars correspond to the respective standard deviations.

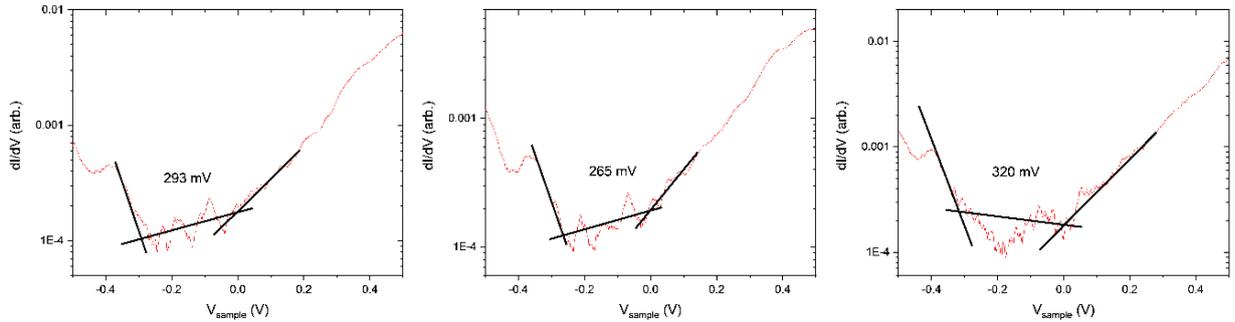

**Figure S7.** 1 SL spectra.

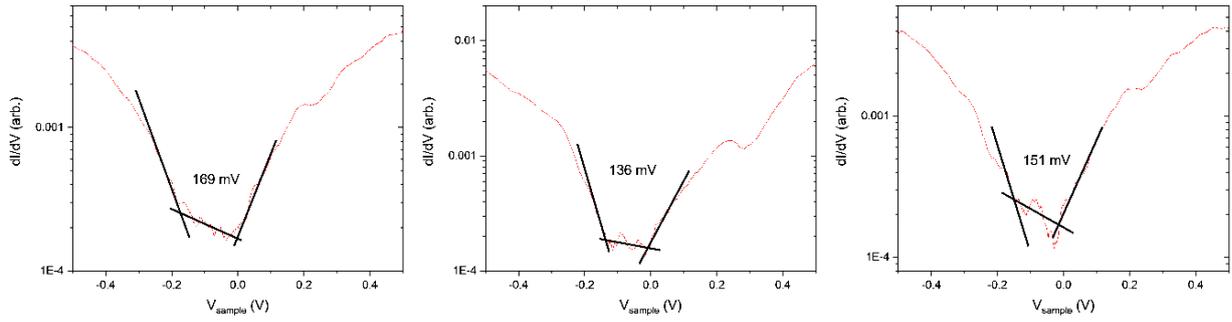

**Figure S8.** 2 SL spectra.

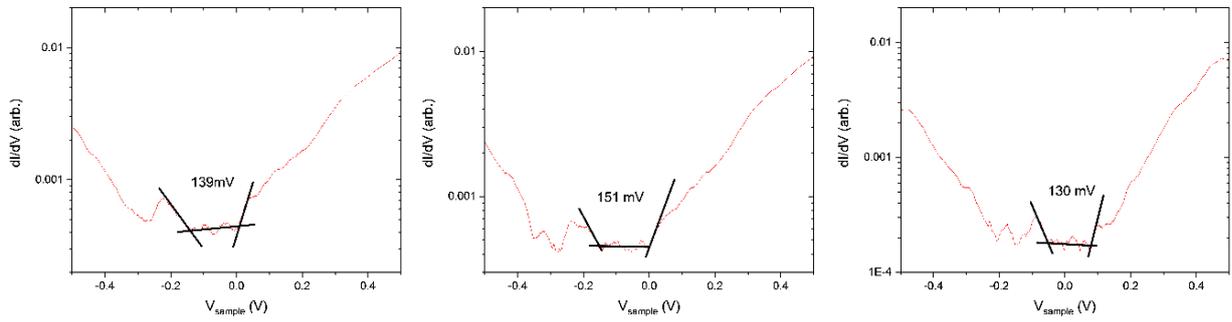

**Figure S9.** 3 SL spectra.

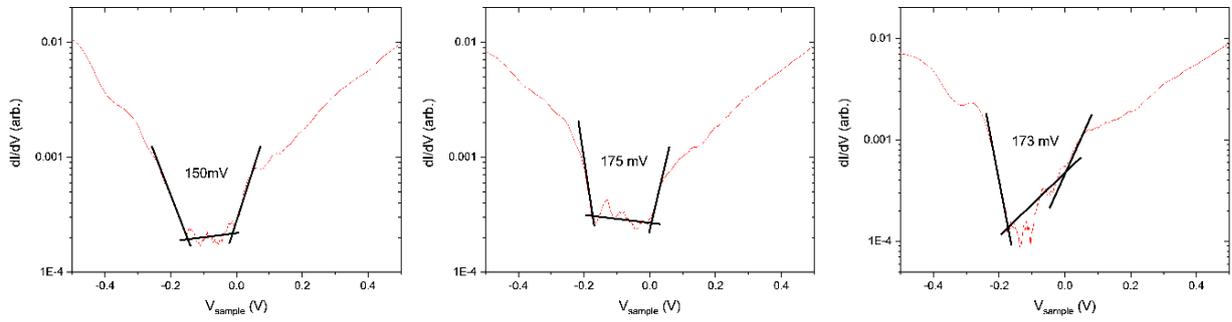

**Figure S10.** 4 SL spectra.

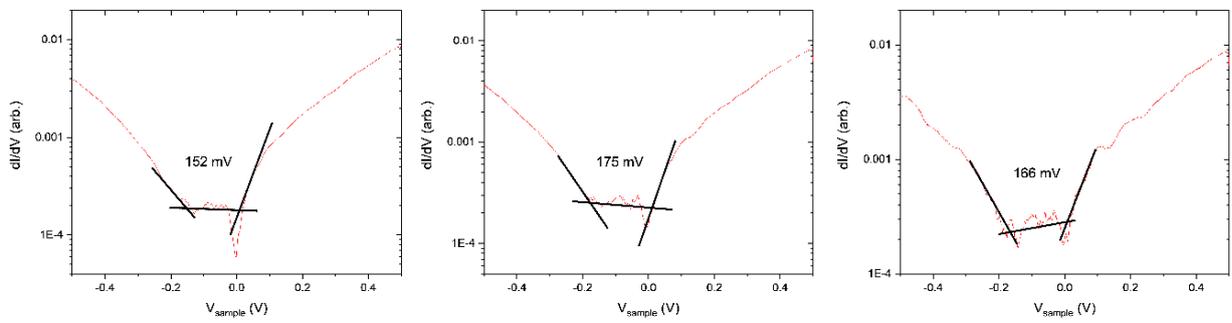

**Figure S11.** 5 SL spectra.

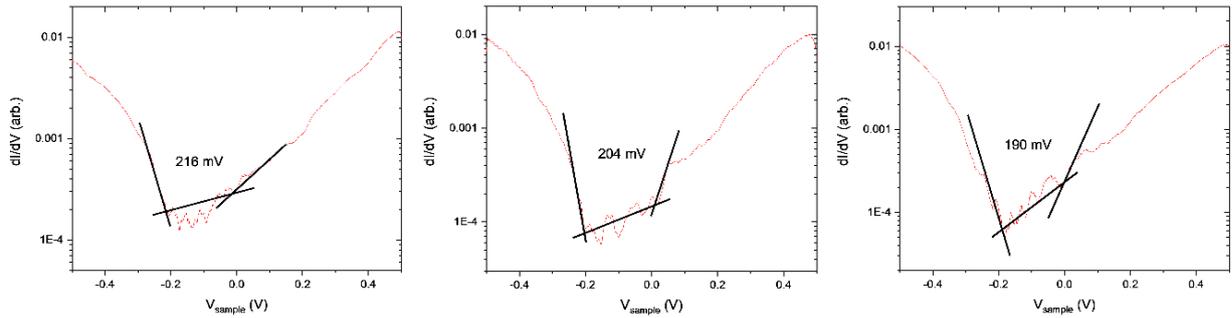

**Figure S12.** 6 SL spectra.

**Supplemental References**